\documentclass{iopart}
\usepackage{iopams}  
\usepackage{graphicx}  
\usepackage{dcolumn}   
\usepackage{bm}        
\usepackage[english]{babel}

\newcommand{\eq}{\begin{equation}}
\newcommand{\eeq}{\end{equation}}
\newcommand{\be}{\begin{equation}}
\newcommand{\ee}{\end{equation}}
\newcommand{\bea}{\begin{equation}}
\newcommand{\eea}{\end{equation}}

\begin{document}

\title{Curvature singularities, tidal forces and the viability of Palatini $f(R)$ gravity}

\author{E Barausse${}^1$, T P Sotiriou${}^{1,2}$ and J C Miller${}^{1,3}$}

\address{${}^1$SISSA, International School for Advanced Studies, 
Via Beirut 2-4, 34014 Trieste, Italy and INFN, Sezione di Trieste}
\address{${}^2$Department of Physics, University of Maryland, College Park, MD 20742-4111, USA}
\address{${}^3$Department of Physics (Astrophysics), University of Oxford, Keble
Road, Oxford OX1 3RH, England}

\eads{\mailto{barausse@sissa.it}, \mailto{sotiriou@umd.edu}, \mailto{miller@sissa.it}}

\begin{abstract}
In a previous paper we showed that static spherically symmetric objects 
which, in the vicinity of their surface, are well-described by a polytropic equation of state with $3/2<\Gamma<2$
exhibit a curvature singularity in Palatini $f(R)$ gravity. We argued that this casts serious doubt on the validity of
Palatini $f(R)$ gravity as a viable alternative to General Relativity. In the present paper we further investigate this 
characteristic of Palatini $f(R)$ gravity in order to clarify its physical interpretation and consequences.
\end{abstract}

\maketitle

\section{Introduction}

According to recent cosmological observations~\cite{obs}, the late time 
evolution of the universe seems to be dominated by 
a cosmological constant or by some unknown form of energy 
(\textit{dark energy}) which mimics the behaviour of a cosmological constant.
The many problems connected with the inclusion of such a constant in the standard 
framework of general-relativistic cosmology~\cite{wein} have led
many authors to consider possible alternative explanations for the cosmological data.
Clearly, since gravity is by far the most important interaction governing the dynamics 
of the universe on large scales, one of these alternatives is to modify the 
theory of gravity itself by changing General Relativity (GR) in some way.

We focus here on one specific generalization of Einstein's theory: Palatini
$f(R)$ gravity~\cite{buch} (see \cite{phd} for a recent review 
of other attempts to generalize GR). As can be found in many textbooks 
(see for example \cite{wald}), Einstein's theory can be derived from
the Einstein--Hilbert action not only by means of the standard metric
variation, but also by taking independent variations with respect to
the metric and the connection. In this approach, known as the \textit{Palatini variational approach},
the metric and the connection are treated as independent quantities, and one 
has to vary the action with respect to both of them in order to obtain the field equations. 
The Riemann tensor $R^\lambda_{\phantom{a}\mu\sigma\nu}$ and the Ricci tensor $R_{\mu\nu}$ 
are defined with respect to the now independent connection  $\Gamma^{\lambda}_{\phantom{a}\mu\nu}$ 
and do not necessarily coincide with the Ricci and Riemann tensors of the metric $g_{\mu\nu}$. 
Similarly, the Ricci scalar is defined as $R=g^{\mu\nu}R_{\mu\nu}$. 
If the Lagrangian is linear in  $R$ (the Einstein--Hilbert action), 
variation with respect to the independent connection forces it to reduce to 
the Levi--Civita connection of the metric, whereas 
variation with respect to the metric gives the standard Einstein equations. 
Therefore, in the case of the Einstein--Hilbert action the outcome of Palatini variation is standard GR. 
However, clearly Einstein's theory is no longer recovered for a generic action
 \be
\label{action}
S=\frac{1}{16\,\pi}\int d^4 x\sqrt{-g}f(R)+ S_M(g_{\mu\nu},\psi),
\ee
 where $f(R)$ is a function of $R$, $g$ is the determinant of the metric
$g_{\mu\nu}$, $S_M$ is the matter action and $\psi$ collectively
denotes the matter fields. (In this equation, as well as in the rest of this paper, 
we are using units in which $c = G = 1$). 
The resulting theory is then what is known as $f(R)$ gravity in the Palatini 
formalism or simply Palatini $f(R)$ gravity. It is easy to see that 
independent variation of the action \eref{action} with respect to the metric and 
the connection gives
\begin{eqnarray}
\label{field1}
&F(R) R_{\mu\nu}-\frac{1}{2}f(R)g_{\mu\nu}=8\,\pi \, T_{\mu\nu},\\
\label{field2}
&\nabla_\sigma[\sqrt{-g} F(R)g^{\mu\nu}]=0,
\end{eqnarray}
where $F(R)=\partial f/\partial R$, $T_{\mu\nu}\equiv
-2(-g)^{-1/2}\delta S_{M}/\delta g^{\mu\nu}$ is the usual
stress-energy tensor of the matter and $\nabla_\mu$ is the covariant
derivative with respect to the connection $\Gamma^{\lambda}_{\phantom{a}\mu\nu}$.  
Note that a crucial assumption has been made in order 
to derive \eref{field1} and \eref{field2}: the matter 
action has been taken to be independent of the connection $\Gamma^{\lambda}_{\phantom{a}\mu\nu}$
[see \eref{action}]. This assumption is physically meaningful because 
it implies that the connection which defines parallel transport, and therefore the covariant derivative of matter fields, 
is the Levi-Civita connection of the metric. This demotes the independent connection to the role of an auxiliary field~\cite{sotlib,sot1,sot2}. 
Additionally, under this assumption, the Levi-Civita connection becomes the one with respect to 
which the matter stress-energy tensor is conserved \cite{koivisto} (which implies, in particular, that test
particles follow geodesics of the metric $g_{\mu\nu}$).
In order to restore the geometrical nature of the independent connection, one has to allow it to couple to the matter. 
This leads to metric-affine $f(R)$ gravity~\cite{sotlib}, which is a different theory
with enriched phenomenology~\cite{sot1, sot2}.

Specific choices for the function $f(R)$ in the action \eref{action} have been shown 
to lead to models of Palatini $f(R)$ gravity which might be able to address 
dark-energy problems~\cite{vollick}. There is now an extensive literature on the  
cosmological aspects of such models and discussing their consistency with cosmological 
~\cite{palcosm} and Solar System constraints~\cite{newt,olmonewt}. 
In a previous paper~\cite{nogopal} we focused on the less well-studied issue 
of finding consistent solutions for static spherically-symmetric matter 
configurations when $f(R)\ne R$. 
In order to be able to treat the field equations analytically, we assumed a polytropic
equation of state (EOS) for the matter, \textit{i.e.}
\eq\label{eq:eos}
p=\kappa\rho_0^\Gamma\,,\quad\Gamma>3/2
\eeq
($p$ and $\rho_0$ are the pressure and the rest-mass density, 
while $\kappa$ and $\Gamma$ are constants). This is  
a very common and useful choice for making simplified calculations both in GR and in Newtonian theory~\cite{polytropes}.
We found that for a  polytropic index in the range $3/2<\Gamma<2$ there exist \textit{no}  static and 
spherically-symmetric regular solutions to the field equations, because curvature singularities 
unavoidably arise at the surface. 

There are four points that ought to be stressed about this result:
\begin{enumerate}
\item It holds also for any EOS which can be approximated, near to the surface,
by a polytrope with $3/2<\Gamma<2$.
\item It is independent of the functional form of $f(R)$ (with the exception 
of some very specific cases, including standard GR) \cite{nogopal}. 
As such, it is applicable not only to specific models, 
but it reveals a generic aspect of Palatini $f(R)$ gravity as a class of theories.
\item The singularities appearing are true curvature singularities 
and not coordinate singularities, \textit{i.e.} the curvature invariants of the metric diverge.
\item Apart from the assumptions already listed, concerning symmetries and EOS, no other 
assumption or approximation has been used. As such, the result applies in all regimes 
ranging from Newtonian weak field to strong gravity.
\end{enumerate}

As noted in~\cite{nogopal}, these results cast some serious doubt on the viability of
Palatini $f(R)$ gravity. In the next section, after briefly reviewing the arguments 
justifying this claim, we further analyze the situation by considering gedanken experiments as a 
powerful tool to investigate the completeness of the theory (section \ref{gedanken}). 
In section \ref{tidal_calc}, we calculate the tidal forces exerted due to the presence of the surface singularities, 
and show that the lengthscale on which they arise is much larger than the lengthscale on which the fluid approximation
breaks down, unless one considers very compact configurations \textit{and} a very special form for the function
$f(R)$ in which one cancels by hand several terms generically expected 
to be present in a cosmological scenario.
(These two hypotheses were assumed in a restricted version of the 
calculation performed in \cite{finns}).
In section \ref{nature}, we then discuss the physical and mathematical 
nature of the problem. This analysis reveals that the presence of the singularities 
is not specifically related to the fluid description of matter, but rather is a feature of 
the differential structure of the equations of the theory and would, in general, become 
even more acute if the fluid approximation were to be abandoned. 
In the same section we also propose ways to generalize the theory in order to avoid these problems. 
In section \ref{concs} we present our conclusions.

\section{Surface singularities and viability}

\subsection{Gedanken experiments and incompleteness}
\label{gedanken}

Clearly, a polytropic EOS is too idealized to give a detailed description for a matter 
configuration resembling an astrophysical star. However, this does 
not at all make polytropes 
physically irrelevant. On the contrary, as well as being widely used 
in GR and in Newtonian theory~\cite{polytropes},  
there are at least two physical matter configurations which are \textit{exactly} described by a $\Gamma=5/3$ polytrope: 
a monatomic isentropic gas and a degenerate non-relativistic electron gas. Note that this value of the polytropic index 
lies well within the range $3/2<\Gamma<2$, for which surface singularities have been shown to appear, 
and so Palatini $f(R)$ gravity does not allow a physical solution
for these configurations (a solution which is singular at the surface  
should be discarded as unphysical). One might, therefore, discard Palatini $f(R)$ gravity 
as a viable alternative to GR already on the basis of
such \textit{gedanken} experiments. Alternatively, one must at least accept that the theory, as it
stands, is \textit{incomplete}, being incapable of describing configurations, 
such as a cloud of monatomic gas, which are
well-described even by Newtonian gravity. Note that this means in particular
that Palatini $f(R)$ gravity does not even reproduce the Newtonian limit in these cases!  

It should be stressed that although the fluid description of matter does indeed conceal
information about the microphysics of the system, this is by no means the cause of the problem discussed here, nor will abandoning the fluid approximation solve the problem, as we will show in section~\ref{nature}. 
On the contrary, one naturally expects that systems such as a monatomic isentropic 
gas or a degenerate electron gas should be describable by a theory of gravity without 
resorting to a statistical description. In our opinion, the inability 
of a theory to provide a classical treatment of macroscopic systems 
without a precise microphysical description is already a very serious shortcoming. This problem does not
arise in standard GR.

\subsection{Stars and tidal forces}
\label{tidal_calc}

In this section we calculate the tidal forces arising due to the presence of the surface singularities which we discovered in 
\cite{nogopal}.
A version of this calculation for a particular restricted form of 
$f(R)$ was performed by Kainulainen \textit{et al.}~\cite{finns},
who found that the lengthscale on which the tidal forces diverge due to the curvature singularity 
was shorter than the mean free path (MFP) in that case, and concluded  
that the system was not then well-described using the fluid 
approximation.
We will now show that while this is correct in the particular case which they considered,
that is a very special one and is not representative of the general situation. Reference~\cite{finns} considered in fact
 the case of a neutron star with $f(R)=R-\mu^4/R$ (where $\mu^2\sim \Lambda$, $\Lambda$ being the value of the cosmological
constant as inferred from cosmological observations).
Although $f(R)=R-\mu^4/R$ can be used to obtain the accelerated expansion of the Universe without resorting to Dark
Energy or a cosmological constant, there is no basic principle from which to derive this functional form, and in 
order to justify it one has to invoke phenomenological arguments based on a series expansion of the unknown
$f(R)$ coming from a consistent high energy theory. As such, there is no reason 
to exclude the presence of quadratic or cubic terms, and indeed the observational 
limits on these terms coming from solar system tests are very loose~\cite{tomo}.
We will show that if one takes $f(R)=R-\mu^4/R+\varepsilon R^2$ 
even with $\varepsilon$ being 
orders of magnitude smaller than the maximum allowed by the solar 
system constraints, the lengthscale 
on which the tidal forces diverge is \textit{much} larger than the
MFP, even in the case of neutron stars. Incidentally, this was expected because we have already shown in
\cite{nogopal} how important the effect of such a tiny  $\varepsilon$ can be in neutron star interiors.
However, even if one cancels \textit{by hand} all of the quadratic and 
cubic terms from the function $f(R)$, thus giving precisely
$f(R)=R-\mu^4/R$, the result of \cite{finns} still does not 
apply for sufficiently diffuse systems, 
where the lengthscale on which the tidal forces diverge is anyway \textit{much} larger than the
MFP.

Let us first recall the notation and the main results of \cite{nogopal}. In particular, 
we write the static spherically symmetric metric as
\begin{equation}
\label{metric}
        ds^2 \equiv -e^{A(r)}{\rm d}t^2 + e^{B(r)}{\rm d}r^2 + r^2{\rm d}\Omega^2,
\end{equation}
and denote the pressure, energy density and stress energy-tensor of the fluid by $p$, $\rho$
and $T^{\mu\nu}$. Also, we define $F(R)\equiv\partial f/ \partial R$ and  ${\cal C}\equiv{dF}/{dp}\,(p+\rho)$,
and use a ``prime'' to denote derivatives with respect to the radial coordinate $r$.
We recall that taking the trace of (\ref{field1}), one gets 
\be
\label{contract}
F(R) R-2f(R)=8\,\pi\,T,
\ee
which is an algebraic equation relating $R$ and $T$ for a given $f(R)$ \cite{sotplb}. 
In particular, equation \eref{contract} implies that in the exterior $R$ is constant 
[\textit{i.e.,} $R=R_0$ with $F(R_0) R_0-2f(R_0)=0$]. Similarly, we denote the exterior
values of $f$ and $F$ with a ``zero'': $f_0\equiv f(R_0)$ and $F_0\equiv F(R_0)$.

Solving (\ref{field2}) for the connection and inserting the resulting expression into (\ref{field1}),
it is possible to rewrite the field equations in a more familiar form:
\begin{eqnarray}
\label{eq:field}
\widetilde{G}_{\mu \nu} &= \frac{8\pi}{F}T_{\mu \nu}- \frac{1}{2}g_{\mu \nu} 
                        \left(R - \frac{f}{F} \right) + \frac{1}{F} \left(
			\widetilde{\nabla}_{\mu} \widetilde{\nabla}_{\nu}
			- g_{\mu \nu} \widetilde{\Box}
		\right) F\nonumber\\
 &-\frac{3}{2}\frac{1}{F^2} \left(
			(\widetilde{\nabla}_{\mu}F)(\widetilde{\nabla}_{\nu}F)
			- \frac{1}{2}g_{\mu \nu} (\widetilde{\nabla}F)^2
		\right),
\end{eqnarray}
where $\widetilde{\nabla}_{\mu}$ and $\widetilde{G}_{\mu \nu}$ are the covariant derivative and Einstein tensor built with 
the Levi-Civita connection of $g_{\mu\nu}$, $\widetilde{\Box}\equiv g^{\mu\nu}\widetilde{\nabla}_{\mu}\widetilde{\nabla}_{\nu}$,
and $F$, $f$ and $R$ are expressed as functions of $T$ using (\ref{contract}). 
Because  \eref{eq:field} reduces  in vacuum  basically to GR with a cosmological constant, it is not
surprising that inserting the ansatz~\eref{metric} into it gives 
that the exterior solution must be the well-known Schwarzschild-DeSitter metric (\textit{i.e.,} Birkhoff's
theorem holds also in  Palatini $f(R)$ gravity). This solution is characterized by two parameters: the total mass
$m_{\rm tot}$, which is fixed by matching with the interior solution imposing continuity at the surface, and the 
cosmological constant, which turns out to be $\Lambda=R_0/4$. Solving \eref{eq:field} in the interior,
it can be shown~\cite{nogopal} that the radial derivative $F'$ is zero at the surface ($r=r_{\rm out}$), while 
\be
\label{eq:F''}
 F''(r_{\rm out})=\frac {\left(R_0 r_{\rm out}^3-8 m_{\rm tot}\right)
{\cal C}'}{8 r_{\rm out} (r_{\rm out}-2 m_{\rm tot})}\,.
\ee
and
\begin{equation}
\label{eq:mprime}
 m_{\rm tot}'(r_{\rm out})=\frac{2 F_0 R_0 r_{\rm out}^2+\left(r_{\rm
out}^3 R_0-8 m_{\rm tot}\right) {\cal C}'}{16 F_0}\;.
 \end{equation}
For $3/2<\Gamma<2$, it can be checked that ${\cal C}'\to\infty$ as the surface is
approached, thus driving to infinity $F''$, $m_{\rm tot}'$ and, more importantly, 
the Riemann tensor of the metric, $\widetilde{R}_{\mu\nu\sigma\lambda}$, and curvature invariants, such
as $\widetilde{R}$ or $\widetilde{R}^{\mu\nu\sigma\lambda}\widetilde{R}_{\mu\nu\sigma\lambda}$.

We will now proceed to calculate in detail the tidal force 
experienced, because of this curvature singularity, by a body 
falling radially into a polytropic sphere with $\Gamma=5/3$, as soon as it crosses the surface.
Our conclusions apply unchanged also to bodies moving 
on different orbits, \textit{e.g.} circular ones 
just below the surface.
Throughout the calculation, we use units in which $M_\odot=1$, as well as $G=c=1$.
If we consider the separation vector $\boldsymbol{\eta}=\eta^r\partial/\partial r$, 
the tidal acceleration in the radial direction is given by the geodesic deviation equation:
\begin{equation}
\frac{D^2 \eta^r}{D\tau^2}=R^r_{\phantom{r}tt r}(u^t)^2\eta^r=
-\frac{1}{4}e^{A-B}(A'^2-A'B'+2 A'')(u^t)^2\eta^r\;,
\end{equation}
where $\tau$ is the proper time and $D/D\tau$ is the total covariant derivative with respect to it.
Using (8)-(11) of \cite{nogopal} and Mathematica~\cite{mathematica}, it is easy to show that the combination 
$A'^2-A'B'+2 A''$ appearing in this equation depends linearly on $F''$:
\eq
A'^2-A'B'+2 A''= c_0+c_1 F''\;,
\eeq
where 
\begin{eqnarray}
c_0&=\Big\{16 F^4+40 r F' F^3+52 r^2 F'^2 F^2\nonumber\\&+16 e^{2
   B} \pi  r^2 \left[(f+16 \pi  p-F R) r^2+2
   F\right] (p+\rho) F^2\nonumber\\&+24 r^3 F'^3 F-2
   e^{B} \left(2 F+r F'\right) [3 (f+12 \pi 
   p-4 \pi  \rho) F' r^3\nonumber\\&-F \left(r \left(r
   \left(f'+2 R F'\right)+8 \pi  \left(p+\rho+2 r
   p'\right)\right)-4 F'\right) r\nonumber\\&+F^2 \left(R'
   r^3+4\right)] F+3 r^4 F'^4\Big\}/[r^2 F^2 \left(2
   F+r F'\right)^2]\;,\label{eq:c_0}
\end{eqnarray}
\begin{equation}\label{eq:c_1}
c_1=-\frac{4}{2 F+r
   F'}\;.
\end{equation}
Note that both $c_0$ and $c_1$ are finite at the surface, whereas $F''$ diverges, as already mentioned. 
Keeping therefore only the divergent term $c_1 F''$, the ratio between the accelerations in the Palatini (``singular'') 
and GR cases is
\eq
\left\vert\frac{a_{\rm sing}}{a_{\rm GR}}\right\vert\approx \frac{\vert c_1 F''\vert r_{\rm out}^2(r_{\rm out}-2m_{\rm tot})}{8 m_{\rm tot}}
\eeq
Using now the fact that $R_0$ must be $\ll1$ 
in our units in order to match the cosmological accelerated expansion (one needs to have $R_0=4 \Lambda=12\Omega_{\Lambda}H_0^2\sim10^{-45}$), \eref{eq:F''} gives $F''\approx-m_{\rm tot}{\cal C}'/[r_{\rm out}(r_{\rm out}-2m_{\rm tot})]$ and therefore
\eq
\left\vert\frac{a_{\rm sing}}{a_{\rm GR}}\right\vert\approx \frac{\vert c_1 {\cal C}'\vert r_{\rm out}}{8}\;.
\eeq
The derivative of $\cal C$ with respect to $r$ can easily be calculated from the definition ${\cal C}\equiv{dF}/{dp}\,(p+\rho)$: 
using the chain rule, the Euler equation $p'=-A'(p+\rho)/2$ and the fact that the trace of the stress energy tensor
for a perfect fluid is $T=3p-\rho$, one has
\begin{eqnarray}
{\cal C}'=\frac{d{\cal C}}{dp}p'&=-\left[\frac{d^2F}{dp^2}(p+\rho)^2+\frac{dF}{dp}\left(1+\frac{d\rho}{dp}\right)(p+\rho)\right]\frac{A}2'\nonumber\\&=
-\frac{A'}{2}\Bigg\{{\cal C}+\frac{dF}{dR}\frac{dR}{dT}\frac{dT}{d\rho}\left(\frac{d\rho}{dp}\right)^2(p+\rho)+
(p+\rho)^2\times\nonumber\\&\quad\Bigg[\frac{dF}{dR}\frac{dR}{dT}\left(-\frac{d^2\rho}{dp^2}\right)+\frac{dF}{dR}\frac{d^2R}{dT^2}\left(3-\frac{d\rho}{dp}\right)^2\nonumber\\&\quad
+\frac{d^2F}{dR^2}\left(\frac{dR}{dT}\right)^2\left(3-\frac{d\rho}{dp}\right)^2\Bigg]\Bigg\}.
\end{eqnarray}
Remembering now that ${\cal C}=0$, $d\rho/dp(p+\rho)=0$ and 
$dp/d\rho=0$ 
at the surface for $\Gamma<2$~\cite{nogopal}, one can easily rewrite the above equation keeping only the divergent terms: 
\begin{eqnarray}
{\cal C}'&=\frac{dF}{dR}\frac{dR}{dT}\frac{A'}{2}
\left[(p+\rho)\left(\frac{d\rho}{dp}\right)^2+\frac{d^2\rho}{dp^2}(p+\rho)^2\right]\nonumber\\&+\mbox{terms going to zero at the surface}
\end{eqnarray}
Taking $\Gamma=5/3$ and using $A'\approx2m_{\rm tot}/[r_{\rm out}(r_{\rm out}-2m_{\rm tot})]$ (see (17) of \cite{nogopal}), this equation becomes
\begin{eqnarray}
{\cal C}'&\approx\frac{dF}{dR}\frac{dR}{dT}\frac{A'}{2}
\left[(p+\rho)\left(\frac{d\rho}{dp}\right)^2+\frac{d^2\rho}{dp^2}(p+\rho)^2\right]\nonumber\\&\approx-8\pi\frac{dF}{dR}\frac{3m_{\rm tot}}{25r_{\rm out}(r_{\rm out}-2m_{\rm tot})\kappa^2}\left(\frac{p}{\kappa}\right)^{-1/5}\;,
\end{eqnarray}
where, in order to pass from the first to the second line, we have used the fact that $dR/dT\approx-8\pi$ close to the surface but at a finite distance below it, for a generic function $f(R)=R-\mu^4/R+\varepsilon R^2$. 
To see this, one can solve \eref{contract} and obtain
$R=-4\pi T\pm(3\mu^4+16\pi^2T^2)^{1/2}$. Choosing the positive sign in order to have a positive cosmological constant in vacuum, one has
$\mu^2=R_0/\sqrt{3}\sim10^{-45}$. Then, even very close to the surface, one has $\vert T\vert\gg \mu^2$ and $R\approx -8\pi T$.

Integrating the Euler equation $p'=-A'(p+\rho)/2$ just below the surface 
one gets
\eq\label{eq:p}
p\approx\left(\frac25\right)^{5/2}\left[\frac{m_{\rm tot}}{r_{\rm out}(r_{\rm out}-2m_{\rm tot})}\right]^{5/2}(r_{\rm out}-r)^{5/2}\kappa^{-3/2}
\eeq
hence 
\eq\label{eq:rho}
\rho\approx\left(\frac25\right)^{3/2}\left[\frac{m_{\rm tot}}{r_{\rm out}(r_{\rm out}-2m_{\rm tot})}\right]^{3/2}(r_{\rm out}-r)^{3/2}\kappa^{-3/2}
\eeq
Therefore,
\eq\label{eq:tidal_forces}
\left\vert\frac{a_{\rm sing}}{a_{\rm GR}}\right\vert\approx\frac{3\sqrt{5}}{25\sqrt{2}}\pi \left\vert c_1 \frac{dF}{dR}\right\vert\frac{\sqrt{m_{\rm tot}\, r_{\rm out}}}{\sqrt{r_{\rm out}-2 m_{\rm tot}}}\frac{\kappa^{-3/2}}{\sqrt{r_{\rm out}-r}}
\eeq

To calculate the ratio given by \eref{eq:tidal_forces}, let us first consider the general case
$f(R)=R-\mu^4/R+\varepsilon R^2$. We stress again that one generically expects 
the presence of the term $\varepsilon R^2$, because there is no first principle
from which to derive the functional form of $f(R)$, and one has to think of it as the series expansion of an unknown
$f(R)$ coming from a consistent high-energy theory of gravity.
As can easily be seen from \eref{contract}, the quadratic term does not influence 
the vacuum value $R_0$ of the curvature scalar, which acts as the effective 
cosmological constant. Basically for this reason, the quadratic term is essentially unconstrained by cosmological data and
solar system tests only allow weak constraints to be placed on it~\cite{tomo}. Taking now $\varepsilon\sim0.1$ in our units 
(a value several orders of magnitude smaller than the upper limit 
coming from solar system tests~\cite{tomo}), 
just below the surface we have $dF/dR\approx2\varepsilon\approx 0.2$ and $c_1\approx -2/F_0\approx-3/2$ (because $F'\sim0$ near to the surface). From 
\eref{eq:tidal_forces}, one then obtains
\eq\label{eq:tidal_forces_quadratic}
\left\vert\frac{a_{\rm sing}}{a_{\rm GR}}\right\vert\approx0.2\frac{\sqrt{m_{\rm tot}\, r_{\rm out}}}{\sqrt{r_{\rm out}-2 m_{\rm tot}}}\frac{\kappa^{-3/2}}{\sqrt{r_{\rm out}-r}}\;.
\eeq 
In the case of a neutron star as modelled with a polytropic EOS  ($\kappa\approx 4$, $r_{\rm out}\approx 10$ and $m_{\rm tot}\approx 1$), one therefore has
\eq\label{eq:tidal_forces_quadratic_NS}
\left\vert\frac{a_{\rm sing}}{a_{\rm GR}}\right\vert\approx 0.02(r_{\rm out}-r)^{-1/2}\;,
\eeq
and the ratio $\vert a_{\rm sing}/a_{\rm GR}\vert$ is large 
at distances below the surface at which the fluid approximation is certainly valid. For instance, 
$\vert a_{\rm sing}/a_{\rm GR}\vert\sim 20$ for $r_{\rm out}-r\sim 10^{-6}\sim1.5$ mm, 
$\vert a_{\rm sing}/a_{\rm GR}\vert\sim 600$ for $r_{\rm out}-r\sim 10^{-9}\sim1.5$ $\mu$m,
$\vert a_{\rm sing}/a_{\rm GR}\vert\sim 2 \times10^4$ for $r_{\rm out}-r\sim 10^{-12}\sim 1.5$ nm.
Note also that the ratio $\vert a_{\rm sing}/a_{\rm GR}\vert$ scales proportionally with the value of $\varepsilon$,
which we have taken, as already mentioned, to be several orders of 
magnitude smaller than the upper limits coming from solar system 
tests~\cite{tomo}.

Let us now consider instead the case $f(R)=R-\mu^4/R$, as used in \cite{finns}.
First, we need to evaluate $c_1$. Noting that $c_1\to-2/F_0=-3/2$ 
as $r\to r_{\rm out}$, because $F'=0$ at the surface~\cite{nogopal}, 
we have $c_1\sim1$ just below the surface. 
In order to see what happens instead at a finite distance below the surface, note first that 
$dF/dR\approx-2R_0^2/(3R^3)$. As already mentioned, solving \eref{contract} 
and imposing that the cosmological constant $\Lambda=R_0/4$ in vacuum is positive, 
one gets $R=-4\pi T+(3\mu^4+16\pi^2T^2)^{1/2}\approx8\pi\rho$ for $\rho\gg R_0$.
We can then write $dF/dR\approx-2/[3(8\pi)^3]R_0^2/\rho^3$ for $\rho\gg R_0$, and therefore $F'=(dF/dR)\,(dR/dT)\,(dT/d\rho)\,\rho'\sim\rho'\,(R_0/\rho)^3/R_0$. 
Because $R_0\sim10^{-45}$, it is clear $rF'\ll F$ even at finite distances below
the surface. From \eref{eq:c_1} it then follows that $c_1\approx -2/F_0=-3/2$ 
also at finite distances below the surface. 

Let us now evaluate \eref{eq:tidal_forces} for $\rho\gg R_0$: using
$c_1\approx -3/2$ and $dF/dR\approx-2/[3(8\pi)^3]R_0^2/\rho^3$, it becomes 
\begin{equation}\label{eq:tidal_forces1}
\left\vert\frac{a_{\rm sing}}{a_{\rm GR}}\right\vert\approx\frac{3\pi}{25(8\pi)^3}\left(\frac52\right)^5 R_0^2\kappa^3 m_{\rm tot}^{-4} {r_{\rm out}}^{5}(r_{\rm out}-2 m_{\rm tot})^{4}(r_{\rm out}-r)^{-5}.
\end{equation}
It is therefore clear that tidal forces become increasingly more important, even in the particular case $f(R)=R-\mu^4/R$, for
spheres with larger radius. As such, even for this particular form of $f(R)$, the lengthscale on which the tidal forces
diverge is much larger than the lengthscale on which the fluid approximation is valid, if one considers sufficiently diffuse
systems: some examples are worked out in the Appendix.

In conclusion, we have shown that the fluid approximation is still valid on the scale
at which the tidal forces diverge just below the surface of a polytropic sphere
in the case of the generic functions $f(R)$ likely to arise
in practice in a cosmological scenario. 
Even in the special case considered by Kainulainen \textit{et 
al.}~\cite{finns}, this continues to hold
for configurations which are sufficiently diffuse.

\section{Physical and mathematical nature of the problem}
\label{nature}

\subsection{Differential structure and cumulativity}
\label{root_sec}

It is clear from the above that the nature of the problem discussed here does not lie in the fluid approximation or in the specifics of the approach followed in \cite{nogopal}, but is related to intrinsic characteristics of Palatini $f(R)$ gravity. These concern
the differential structure of the action~\eref{action} and the resulting field equations.

We recall that the Lagrangian of the action (\ref{action}) is an algebraic function of $R=g^{\mu\nu}R_{\mu\nu}$ and that $R_{\mu\nu}$ is constructed from the independent connection $\Gamma^\lambda_{\phantom{a}\mu\nu}$. In more detail,
\be
\label{riemann}
R^\mu_{\phantom{a}\nu\sigma\lambda}=-\partial_\lambda\Gamma^\mu_{\phantom{a}\nu\sigma}+\partial_\sigma\Gamma^\mu_{\phantom{a}\nu\lambda}+\Gamma^\mu_{\phantom{a}\alpha\sigma}\Gamma^\alpha_{\phantom{a}\nu\lambda}-\Gamma^\mu_{\phantom{a}\alpha\lambda}\Gamma^\alpha_{\phantom{a}\nu\sigma}\, ,
\ee
and contracting the first and the third index one gets (see \cite{phd,sotlib} for further details)
\begin{equation}
\label{ricci}
R_{\mu\nu} =R^\lambda_{\phantom{a}\mu\lambda\nu}=\partial_\lambda \Gamma^\lambda_{\phantom{a}\mu\nu}-\partial_\nu \Gamma^\lambda_{\phantom{a}\mu\lambda}+\Gamma^\lambda_{\phantom{a}\sigma\lambda}\Gamma^\sigma_{\phantom{a}\mu\nu}-\Gamma^\lambda_{\phantom{a}\sigma\nu}\Gamma^{\sigma}_{\phantom{a}\mu\lambda}.
\end{equation}
It follows from this that $f(R)$ has no \textit{a priori} dependence on derivatives of the metric. Also, $R$ depends only linearly on the first derivatives of the connection {\it i.e.,} at least in the case where $f$ is linear in $R$ (which leads to GR), there are no $\partial \Gamma \partial \Gamma$ terms (indices suppressed) as there would usually be in a field theory!
One might expect that allowing $f$ to be non-linear in $R$ would introduce $\partial \Gamma \partial \Gamma$ terms and solve this last problem, but we will see shortly that this is not the case.
Note also that, since the metric has no {\it a priori} relation with the connection, one is dealing with a field theory with two independent fields, and so one cannot argue that having no quadratic terms in the connection is expected because the connection already includes derivatives of the metric. 

This lack of dynamics in the action is also mirrored in the field equations (\ref{field1}) and (\ref{field2}). Variation with respect to the metric leads to (\ref{field1}), which includes no derivatives of the metric. As already mentioned, contraction of (\ref{field1}) gives \eref{contract}, which algebraically relates $R$ and $T$ for a given $f(R)$. Variation with respect to the connection leads to (\ref{field2}), after some integration by parts to ``free'' the connection. For a linear function $f(R)$, this equation is just the definition of the Levi-Civita connection. When $f(R)$ is non-linear, instead, (\ref{field2}) seems to include second derivatives of the connection. However, this is misleading because $R$ can be completely eliminated in favour of $T$ by using (\ref{contract}) and, therefore, (\ref{field2}) can be trivially solved to give the connection as a function of the metric and the matter fields. 
As already mentioned, following these steps one can completely eliminate the connection in favour of the metric and the matter fields, and turn (\ref{field1}) and (\ref{field2}) into the single-field representation~\eref{eq:field}. This representation of 
the theory is more convenient and more familiar for discussing the dynamics. It also highlights once more that the metric fully describes the geometry, which is indeed pseudo-Riemannian, and that the independent connection is just an auxiliary field \cite{sotlib,sot1,sot2}. 

It is also interesting to note that one could introduce an auxiliary scalar $\phi=F$ and re-write 
\eref{eq:field} as
 \begin{eqnarray}
\label{eq:field2}
\widetilde{G}_{\mu \nu} &= \frac{8\pi}{\phi}T_{\mu \nu}- \frac{1}{2}g_{\mu \nu} 
                        \left(R - \frac{f}{\phi} \right) + \frac{1}{\phi} \left(
			\widetilde{\nabla}_{\mu} \widetilde{\nabla}_{\nu}
			- g_{\mu \nu} \widetilde{\Box}
		\right) \phi\nonumber\\
&-\frac{3}{2}\frac{1}{\phi^2} \left(
			(\widetilde{\nabla}_{\mu}\phi)(\widetilde{\nabla}_{\nu}\phi)
			- \frac{1}{2}g_{\mu \nu} (\widetilde{\nabla}\phi)^2
		\right),		
\end{eqnarray}
while, setting $V(\phi)=R\phi-f$, (\ref{contract}) can be re-written as
\be
\label{sc}
2V(\phi)-\phi V'(\phi)=8\,\pi\, T.
\ee
Expressions (\ref{eq:field2}) and (\ref{sc}) are the field 
equations 
of a 
Brans--Dicke theory with Brans-Dicke parameter $\omega_0=-3/2$, {\it i.e.}~a theory described by the action
\begin{equation}
\label{palactionH2d0}
S= \frac{1}{16\,\pi}\int d^4 x \sqrt{-g} \left(\phi \widetilde{R}+\frac{3}{2\phi}\partial_\mu \phi \partial^\mu \phi-V(\phi)\right) +S_M(g_{\mu\nu}, \psi),
\end{equation}
(see also \cite{sot1,flanagan,scpal} for more details about the equivalence of Palatini $f(R)$ gravity and $\omega_0=-3/2$ Brans-Dicke theory).

Returning to (\ref{eq:field}), we note that this is a second order partial differential equation in the metric, just as in the case of GR, but that the left hand side includes up to second derivatives of $F$ and consequently of $T$ [$F=F(R)$ and $R=R(T)$]. Usually, the matter action includes derivatives of the matter fields $\psi$ (if the equation of motion of the matter fields is to be  of second order, the matter action has to be quadratic in the first derivatives of the matter fields). Therefore, generically one has $T=T(\partial \psi,\psi)$, implying that (\ref{eq:field}) includes up to third derivatives of the matter fields! 

In GR and in most of the proposed alternatives to it, the field equations include only first derivatives of the matter fields. The higher differential order in the metric with respect to the differential order in the matter fields guarantees that the metric comes as an integral over the matter fields. Therefore, any discontinuities in the matter are ``smoothed out'' and are not inherited by the geometry (cumulativity of gravity). We recall that in general the metric 
is not allowed to become a delta function or a step function (although the latter is allowed if no Dirac deltas are produced 
in the field equations, \textit{i.e.} if the metric, in spite of being discontinuous, is a solution 
of the field equations in the sense of distributions: see for instance 
\cite{poisson}, section 3.7). However, this is clearly not true in Palatini $f(R)$ gravity or in $\omega_0=-3/2$ Brans--Dicke theory, since the differential order of the field equations in the matter fields is actually higher than in the metric, implying that the latter is not necessarily an integral over the matter fields, but can be \textit{algebraically} related to the matter fields and  even to their derivatives.  Because of this, a discontinuity in the matter fields or in their derivatives can lead to unacceptable singularities. A similar behaviour has been demonstrated in the post-Newtonian limit of the theory, where the post-Newtonian metric becomes algebraically dependent on the matter density \cite{olmonewt}.

This unusual differential structure of Palatini $f(R)$ gravity is at the root of the surface singularities discovered in  \cite{nogopal}. The polytropic description of matter was used in \cite{nogopal} only because this made it possible to find analytic solutions and demonstrate the problem without resorting to numerical techniques. In fact, a more detailed description of the matter would make the problem even more acute. To see this, note that in the case of a perfect fluid one has $T_{\mu\nu}=T_{\mu\nu}(\rho,p)$, {\it i.e.} the stress-energy tensor does not include any derivatives, unlike the case of a microscopic description of matter. The fluid approximation actually ``smoothes out'' the matter distribution with respect to the microscopic description. This also explains why no singularities appear for $1<\Gamma<3/2$: these values of $\Gamma$ give a smooth passage from the interior to the exterior. In conclusion: abandoning the fluid approximation would just increase even further the differential order of the field equations in the matter fields and make it easier for singularities to appear.

As a further confirmation that the introduction of microphysics cannot solve the problems caused by the algebraic dependence
of $R$ on $T$ [see \eref{contract}] or, in the equivalent action \eref{palactionH2d0}, by the algebraic dependence
of $\phi$ on $T$ [see \eref{sc}], let us note that this feature of Palatini $f(R)$ gravity introduces corrections to the standard model
of particle physics already at the meV energy scale
(see \cite{flanagan} and \cite{padilla}). Both the calculation of \cite{flanagan} and that of \cite{padilla}
are performed in the Einstein frame. Although the use of the Einstein 
frame has been criticized \cite{vollick_vs_flanagan}\footnote{If the independent connection is allowed to enter the matter action,
the results of \cite{flanagan} and \cite{padilla} will of course cease to hold, 
as pointed out in \cite{vollick_vs_flanagan}. In this case, also the surface singularities that 
we found in \cite{nogopal} may disappear (see also section \ref{weshallovercome}).
However, such a theory would be a \textit{generalization}
of Palatini $f(R)$ gravity [see the action \eref{action}], 
known in the literature as \textit{metric affine $f(R)$ gravity} \cite{sotlib}.}, this frame
is equivalent to the Jordan frame and both are perfectly suitable 
for performing calculations \cite{flanagan}.
However, one should remember that particles in vacuum follow geodesics of the Jordan frame metric, so
this is the metric which becomes approximately Minkowski in the laboratory reference frame. 
This makes the Jordan frame calculation simpler and more transparent than
the one in the Einstein frame. For this reason, and in order to highlight once again the problems caused by the algebraic dependence
of $R$ on $T$, we briefly redo the calculation of \cite{flanagan} and \cite{padilla} 
in the Jordan frame.
Let us first consider the equivalent action \eref{palactionH2d0}
and take the matter to be represented by a scalar field $H$ (\textit{e.g.,} the Higgs boson), the Lagrangian of which reads
\eq\label{higgs}
{\cal L}_{m}= \frac{1}{2\hbar}\left(g^{\mu\nu}\partial_\mu H \partial_\nu H -\frac{m_{\rm H}^2}{\hbar^2} H^2\right)
\eeq
(we recall that we are using units in which $G=c=1$).
The vacuum of the action \eref{palactionH2d0} with \eref{higgs} and $f(R)=R-\mu^4/R$ [which implies $V(\phi)=2\mu^2 (\phi-1)^{1/2}$]
can easily be found to be $H=0$, $\phi=4/3$ [the solution of \eref{sc} with $T=0$] and 
\begin{eqnarray} 
&g_{\mu\nu}dx^\mu dx^\nu=-\left(1- \frac{2M_{_{\rm Earth}}}{r}-\frac{\mu^2 r^2}{4\sqrt{3}}\right)dt^2\nonumber\\
&+dr^2\Big/\left(1- \frac{2M_{_{\rm Earth}}}{r}-\frac{\mu^2 r^2}{4\sqrt{3}}\right)+r^2d\Omega^2\approx\eta_{\mu\nu}dx^\mu dx^\nu 
\end{eqnarray}
(which is indistinguishable from the Minkowski metric for the purposes of a particle physics experiment because $\mu^2\sim\Lambda$ and $r\approx R_{_{\rm Earth}}$). 
One can then expand the action to second order around this vacuum 
(as usual the first order action is identically zero because the field equations are satisfied to zeroth order). However,
it is easy to show that perturbing \eref{sc} one gets $\delta\phi\sim T/\mu^2\sim m_{\rm H}^2 \delta\! H^2/(\hbar^3 \mu^2)$
at energies lower than the Higgs mass ($m_{\rm H} \sim 100-1000$ GeV): replacing this
expression in the action \eref{palactionH2d0} perturbed to second order one immediately gets that the effective Lagrangian for the Higgs scalar is 
\begin{eqnarray}\label{higgs2}
{\cal L}_{m}^{\rm effective}\sim  &\frac{1}{2\hbar}\left(g^{\mu\nu}\partial_\mu\delta\! H \partial_\nu \delta\! H 
-\frac{m_{\rm H}^2}{\hbar^2} \delta\! H^2\right)\nonumber\\&\times \left[1+
\frac{m_{\rm H}^2 \delta\! H^2}{\mu^2\hbar^3}+\frac{m_{\rm H}^2 (\partial \delta\! H)^2}{\mu^4\hbar^3}\right]
\end{eqnarray}
at energies $k\ll m_{\rm H}$.
At an energy $k=10^{-3}$ eV (corresponding to a lengthscale $L=\hbar/k=2\times 10^{-4}$ m), using the fact that 
$\mu^2\sim\Lambda\sim 1/(H_0^{-1})^2$ (where $H_0^{-1}=4000$ Mpc is the Hubble radius) and $\delta \! H\sim m_{\rm H}$ (because $k\ll m_{\rm H}$)
and remembering that we are using units in which $G=c=1$, it is easy to check that the first correction 
is of the order $m_{\rm H}^2 \delta\! H^2/(\mu^2\hbar^3)\sim (H_0^{-1}/\lambda_{\rm H})^2(m_{\rm H}/M_{\rm P})^2\gg1$, 
where $\lambda_{\rm H}=\hbar/m_{\rm H}\sim2\times 10^{-19}-2\times 10^{-18}$ m is the Compton
length of the Higgs and $M_{\rm P}=\hbar^{1/2}=(\hbar c^5/G)^{1/2}=1.2\times 10^{19}$ GeV is the Planck mass.\footnote{Equivalently,
one can write the first correction as a self-interaction term $m_{\rm H}^4 \delta\! H^4/(\mu^2\hbar^6)$: restoring the dependence on $G$ 
this term becomes $G m_{\rm H}^4 \delta\! H^4/(\mu^2\hbar^6)$. In ``particle physics units'' $\hbar=c=1$, the coupling constant
is dimensionless and is given by $G m_{\rm H}^4/\mu^2\sim(H_0^{-1}/\lambda_{\rm H})^2(m_{\rm H}/M_{\rm P})^2\gg1$.}
Similarly, the second correction is of the order $m_{\rm H}^2 (\partial \delta\! H)^2/(\mu^4\hbar^3)\sim (H_0^{-1}/\lambda_{\rm H})^2(m_{\rm H}/M_{\rm P})^2 (H_0^{-1}/L)^2\gg 1$.

Note that replacing $\delta\phi\sim m_{\rm H}^2 \delta\! H^2/(\hbar^3 \mu^2)$ in \eref{palactionH2d0} 
gives also that the coupling of matter to gravity is described by the interaction Lagrangian 
\eq {\cal L}_{\rm int}\sim  \frac{m_{\rm H}^2 \delta\! H^2}{\hbar^3} \left(\delta g+\frac{\partial^2 \delta g}{\mu^2}\right)\sim 
\frac{m_{\rm H}^2 \delta\! H^2}{\hbar^3} \delta g\left[1+  \left(\frac{H_0^{-1}}{L}\right)^2\right]\,.
\eeq 
It is therefore clear that also the coupling to gravity becomes non-perturbative at microscopic scales.
This is, once again, a consequence of the algebraic dependence of $\phi$ on $T$, encoded in \eref{sc},
and this is in agreement with the singularities that we discuss in this paper. 

\subsection{Overcoming the problem}
\label{weshallovercome}
In section \ref{root_sec} we have traced the root of the 
problem: it lies in the awkward differential structure of the field 
equations, in which the matter field derivatives are of higher order 
than the metric derivatives. This introduces non-cumulative effects 
and makes the metric extremely sensitive to the local characteristics 
of the matter. With this in mind, it is not difficult to propose 
a possible way out. Clearly, one would like to restore the cumulative nature of gravity. This requires the introduction of more dynamics into the gravitational sector of the theory. As an example of how to introduce more dynamics, let us consider a theory described by the action:
 \be
\label{action2}
S=\frac{1}{16\,\pi}\int d^4 x \sqrt{-g}\left(R+a R^{\mu\nu}R_{\mu\nu}\right)+ S_M(g^{\mu\nu},\psi )\,,
\ee
 where $a$ should be chosen so as to have the correct dimensions. The term $ R^{\mu\nu}R_{\mu\nu}$ is quadratic in the derivatives of the connection [see (\ref{ricci})]. This implies that the action (\ref{action2}), even though linear in both $R$ and $R^{\mu\nu}R_{\mu\nu}$, is quadratic in the derivatives of the connection and will not lead to an algebraic equation for the connection, unlike the action (\ref{action}). 
Indeed, the field equations that one derives by varying the action \eref{action2} with respect to the metric and the connection are, respectively,
\begin{eqnarray}
\label{palfg1}
&R_{(\mu\nu)}+2a R^{\sigma}_\mu R_{\sigma\nu}- \frac{1}{2}\left(R+a R^{\sigma\lambda} R_{\sigma\lambda}\right) g_{\mu\nu} =8\pi\, T_{\mu\nu},\\
\label{palfg2}
 &\nabla_\lambda\left[\sqrt{-g}\left(g^{\mu\nu}+2a  R^{\mu\nu}\right)\right] =0,
\end{eqnarray}
and (\ref{palfg2}) cannot be algebraically solved for the connection. 
Also, this theory cannot be re-written as an $\omega_0=-3/2$ 
Brans--Dicke theory. In summary, a theory described by action 
(\ref{action2}) does not seem to be sharing the unwanted 
characteristic of Palatini $f(R)$ gravity: that after eliminating the 
connection, one ends up with the matter field derivatives being of 
higher order than those of the metric. In particular, $R$ cannot be 
expressed as an algebraic function of $T$ though the trace of 
eq.~(\ref{palfg1}), as in the case of eq.~(\ref{contract}), nor can 
the independent connection be algebraically expressed simply in terms 
of the metric and derivatives of the matter fields (therefore 
introducing the higher differential order in the matter fields when 
it is replaced in the field equations). 

Theories with higher order invariants in the action, such as 
$R_{\mu\nu}R^{\mu\nu}$, have recently been considered in the Palatini formalism
in \cite{Li:2007xw}. Clearly, a more detailed analysis of the dynamics of
such theories is needed in order to show in a clear way whether 
they exhibit the problem discussed here or other viability issues. 
Here we have used them solely to demonstrate that it might be possible 
to overcome the issues discussed here by generalizing the action. This 
clarifies the following point: such shortcomings are not generic to 
Palatini variation, but seem to be a specific problem of Palatini 
variation when used with the specific choice of $f(R)$ actions.

\section{Conclusions}
\label{concs}

In this paper, we have discussed in detail the issue, raised in 
\cite{nogopal}, of surface curvature singularities appearing for 
polytropic spheres in Palatini $f(R)$ gravity. Simple gedanken 
experiments lead us to conclude that the presence of these 
singularities casts serious doubts on the viability of the 
gravity theory. 
Concerning the objection, raised in \cite{finns},
that polytropic EOS's are too idealized to allow one to rule out Palatini $f(R)$ gravity, we stress that among the EOS's not giving a regular static
spherically symmetric solution~\cite{nogopal} there are perfectly physical cases such as a degenerate non-relativistic electron gas or an isentropic monatomic gas. Regular solutions
for these configurations exist even in Newtonian mechanics, and we have argued that a theory not providing such solutions should be considered, at best, 
as being incomplete and as being disfavoured for giving viable alternatives to GR. 
We have also presented quantitative results for the 
magnitude of the tidal 
forces exerted just below the surface of polytropic spheres, showing that,
for generic forms of $f(R)$, the lengthscale on which the tidal forces diverge due to 
the curvature singularities is \textit{much} larger than the lengthscale at which the fluid approximation breaks down.
This generalizes the calculation of Kainulainen \textit{et al.} 
\cite{finns}: while their result (that the tidal forces only diverge 
at lengthscales 
on which the fluid approximation is not valid) is correct for their 
particular choice of $f(R)$ and for neutron stars,
we find that it does not hold, even with their choice of $f(R)$, for more 
diffuse configurations and does not hold,
even in the case of neutron stars, for more plausible choices of $f(R)$.
Finally, an analysis of the differential structure of the field equations for the theory has been presented, which sheds light on the origin of the problem, showing that the appearance of singularities is not related to the fluid approximation. On the contrary, abandoning the fluid approximation would make the problem even more acute. The addition of more dynamics to the theory seems to be a potential way out of this difficulty.

\section*{Appendix}

In this appendix we evaluate \eref{eq:tidal_forces1} -- which gives the ratio
$a_{\rm sing}/a_{\rm GR}$ in the special case of $f(R)=R-\mu^4/R$ --
in several contexts. Our calculations will show that even with 
this special choice of $f(R)$,
the fluid approximation is still valid on the scale at which the tidal forces diverge if
the configuration under consideration is sufficiently diffuse. As in section \ref{tidal_calc}, we use units in which
$G=c=M_\odot=1$.

Let us first consider a solar type star with mass $m_{\rm
star}\approx1$ surrounded by a gas cloud with mass $m_{\rm
cloud}\approx 0.01$ and radius $r_{\rm out}\approx10^{14}$ km, composed
of monatomic isentropic gas. This is a perfectly physical
configuration, although possibly not an \textit{astrophysically} fully
realistic one (note however that $r_{\rm out}\approx10^{14}$ km
is approximately the outer radius of the Oort cloud).
The total mass of such a system is
$m_{\rm tot}\approx m_{\rm star}+m_{\rm cloud}\approx1$, and the polytropic
constant of the cloud is $\kappa\approx9\times10^{12}$. From
\eref{eq:tidal_forces1}, one gets
\eq\label{eq:tidal_forces4}
\left\vert\frac{a_{\rm sing}}{a_{\rm GR}}\right\vert\approx 6\times10^{70}(r_{\rm out}-r)^{-5}\;,
\eeq
and the tidal force becomes
comparable to that of GR at a distance below the surface comparable to
$r_{\rm out}$! Taking, for instance, a value of $r_{\rm out}-r\approx r_{\rm out}/10\sim 10^{13}$,
the tidal forces in Palatini $f(R)$ gravity would be 6 orders of magnitude larger than in GR.
At this distance from the surface, the mean distance
between the particles of the fluid is $\ell\approx 1/n^{1/3}\approx(m_{\rm
 p}/\rho)^{1/3}\approx10^{-5}$ (where $n=\rho/m_{\rm p}$ is the number density and $m_{\rm p}\sim 10^{-57}$
is the mass of the proton). An upper limit for
the MFP at this distance from the surface can be calculated assuming a cross section $\sigma\sim (1\mbox{ \AA})^2\approx 5\times10^{-27}$, giving $\ell_{\rm MFP}\sim 1/(n\sigma)\sim m_{\rm p}/
(\rho\sigma)\sim 10^{11}\ll r_{\rm out}-r$ \footnote{This is an upper limit because it assumes
a ``geometrical'' cross section for encounters between the atoms ($1$ \AA\ is approximately the size of a hydrogen atom). However, 
for hydrogen-hydrogen collisions in the lab, $\sigma({\rm HH}) \sim 20$ \AA$^2$~\cite{HH} while, for instance, hydrogen-lithium collisions have cross sections which are about $1200$ \AA$^2$~\cite{lithium}. More importantly, if the fluid is (even partly) ionized, the cross section can be \textit{much} larger, because Coulomb forces are long range (in strongly coupled plasmas it is actually common to have a MFP shorter than the interparticle distance~\cite{plasma_book}). 
}.
Also, note that the average velocity of the particles in the cloud
can be evaluated from $p=nk_{\rm B}T$ ($k_{\rm B}$ and $T$ being Boltzmann's constant
and the temperature) using \eref{eq:p} and \eref{eq:rho}, and is $v_{\rm av} \sim \{m_{\rm tot}(r_{\rm out}-r)/[r_{\rm out}(r_{\rm out}-2 m_{\rm tot})]\}^{1/2}$. For $r_{\rm out}-r\sim r_{\rm out}/10$, one has $v_{\rm av}\sim 4\times10^{-8}$, which is comparable with the virial velocity
$v_{\rm virial} \sim (m_{\rm tot}/r)^{1/2}\sim10^{-7}$, and so the polytropic 
coefficient $\kappa$ needed to support the cloud is plausible.
In conclusion: for this configuration, the lengthscale on which the tidal forces 
in Palatini $f(R)$ gravity are larger than in GR is certainly larger
than the lengthscale on which the fluid approximation is valid, whether this scale is taken to be the mean interparticle
distance or the MFP. 

We would argue, however, that the relevant scale is actually the interparticle distance, because we are
considering here static \textit{equilibrium} configurations. 
A way to understand this point is to consider how one derives the hydrodynamic equations from the Vlasov equation (\textit{i.e.,} the conservation equation
for the phase-space distribution $f(\boldsymbol{x},\boldsymbol{v})$ in the case of a \textit{collisionless} fluid, that is, one with \textit{infinite} MFP; see for instance \cite{vlasov}, paragraph 27):
\eq\label{eq:vlasov}
\frac{\partial f}{\partial t}+\frac{\partial f}{\partial x^i}v^i+\frac{\partial f}{\partial v^i}\frac{F^i}{m}=0\;,
\eeq
where $v^i=dx^i/dt$ is the velocity, $m$ is the mass of the particles and $F^i$ is the force (thought of as dependent only on position and not on velocity).
By integrating over all velocities, one easily obtains the mass conservation equation $\partial_t\rho+\boldsymbol{\nabla}\cdot(\rho\boldsymbol{\bar{v}})=0$, 
where $\rho=m\int f d^3v$ is the density and $\bar{v}^i=m\int  f v^i d^3v/\rho$ is the average (\textit{i.e.,} macroscopic) velocity.
Similarly, one can multiply \eref{eq:vlasov} by $v^i$ and integrate over all velocities. 
If the velocity distribution
is isotropic \footnote{Of course, one may object that if the fluid is collisionless
there is no interaction which can make the velocity distribution isotropic. However, we are interested here
in showing that for equilibrium configurations the MFP has nothing to do, from the conceptual point of view, with the lengthscale at which the fluid approximation breaks down. 
Moreover, one can always think of a tiny interaction between the particles (resulting
in a \textit{huge} MFP) which can make the velocity distribution isotropic in a sufficiently long time (comparable with the \textit{mean free time}). 
From the conceptual point of view,
one can also think of shooting a beam of collisionless particles into a box (or a potential well): the initially focused velocity distribution
will become isotropic due to the small irregularities in the walls of the box or in the gravitational field.
} 
one then gets the Euler equation $(\partial_t+\boldsymbol{\bar{v}}\cdot\boldsymbol{\nabla})\boldsymbol{\bar{v}}=-(\boldsymbol{\nabla}p)/\rho+\boldsymbol{F}/m$,
where one uses the isotropy of the velocity distribution to define the pressure as $p\delta_{ij}=m\int  f (v^i-\bar{v}^i)(v^j-\bar{v}^j) d^3v$.
If the phase-space distribution $f$ is specified, the mass conservation and the Euler equations are clearly a closed system of equations. Therefore, one does not need to consider 
higher order moments of the Vlasov equation, and the system under consideration is a fluid in spite of the MFP being infinite. 
A typical example of this situation is, for instance, that of Dark Matter in a Friedmann-Robertson-Walker universe. 
Since the velocity distribution is isotropic because of the cosmological principle, 
Dark Matter can be treated, at the background level, as being a fluid (\textit{cf.} for instance
(3.10)--(3.12) of \cite{Kolb}). 
Similarly, for a gas trapped in a box, the fluid approximation 
is valid on scales larger than the interparticle distance, whereas the MFP can be infinite if the fluid is non-collisional.
This is indeed the case considered in the textbook derivation of the perfect gas law (see, for instance,
Landau, Lifshitz and Pitaevskii~\cite{textbook2}, chapter 4), where the only necessary hypothesis is the isotropy of the 
velocity distribution. The role of the box is, in our case, played by 
the gravitational potential well.

  Let us now consider a polytropic sphere with $m_{\rm tot}\approx 0.1$ and
  $r_{\rm out}\approx200 R_\odot\approx10^8$ ($R_\odot$ being the
  radius of the Sun). The polytropic constant is then $\kappa\approx
  2\times10^7$, and \eref{eq:tidal_forces1} becomes
  \eq\label{eq:tidal_forces2}
  \left\vert\frac{a_{\rm sing}}{a_{\rm GR}}\right\vert\approx 10^5(r_{\rm out}-r)^{-5}
  \eeq
   At a distance of $\sim 1.5$ km below the surface, therefore, tidal
  forces are $\sim 10^5$ times stronger in Palatini gravity than in GR,
  while the forces in the two cases become comparable at a distance
  $\gtrsim15$ km. Now, from \eref{eq:rho}, at a distance $r_{\rm
  out}-r\approx1.5\mbox{ km}\approx1$ we have
  $\rho\approx10^{-37}\sim10^{8}R_0$. 
  Although $r_{\rm out}-r$ is certainly smaller than the 
  \textit{upper} MFP limit introduced above, at this density the mean distance
  between the particles of the fluid is $\ell\approx 1/n^{1/3}\approx(m_{\rm
  p}/\rho)^{1/3}\approx10^{-7}\sim0.1$ mm.

 The same considerations apply, although marginally, for a polytropic
 sphere with $m_{\rm tot}\approx 1$ and $r_{\rm out}\approx R_\odot
 \approx5\times10^5$. The polytropic constant is then $\kappa\approx
 2\times10^5$, and \eref{eq:tidal_forces1} becomes
 \eq\label{eq:tidal_forces3}
 \left\vert\frac{a_{\rm sing}}{a_{\rm GR}}\right\vert
 \approx 10^{-26}(r_{\rm out}-r)^{-5}\;,
 \eeq
 from which it follows that the difference between the tidal forces 
 becomes important for $r_{\rm
 out}-r\lesssim7\times10^{-6}\sim1$ cm. At this distance below the
 surface the density is $\rho\approx5\times10^{-34}\sim10^{11}R_0$, and
 the mean distance between the fluid particles is $\ell\approx
 1/n^{1/3}\approx(m_{\rm p}/\rho)^{1/3}\approx10^{-8}\sim0.01$
 mm.

\end{document}